\documentclass[preprint,preprintnumbers,amsmath,amssymb]{revtex4}

\usepackage{graphicx}
\usepackage{dcolumn}
\usepackage{bm}
\usepackage{epstopdf}
\usepackage{textcomp}
\usepackage{color}

\begin{document}
\bibliographystyle{biophysj}

\title{Helicase processivity and not the unwinding velocity exhibits universal increase with force}

\author{David L. Pincus, Shaon Chakrabarti
  and D. Thirumalai} \affiliation{Biophysics Program, Institute For
  Physical Science and Technology, University of Maryland, College
  Park, MD 20742}

\date{\today}
\begin{abstract}

Helicases, involved in a number of cellular functions, are motors that translocate along single-stranded nucleic acid and couple the motion to unwinding double-strands of a duplex nucleic acid. The junction between double and single strands creates a barrier to the movement of the helicase, which can be manipulated \textit{in vitro} by applying mechanical forces directly on the nucleic acid strands.  Single molecule experiments have demonstrated that the unwinding velocities of some helicases increase dramatically with increase in the external force, while others show little response.  In contrast, the unwinding processivity always increases when the force increases. The differing responses of the unwinding velocity and processivity to force has lacked explanation. By generalizing a previous model of processive unwinding by helicases, we provide a unified framework for understanding the dependence of velocity and processivity on force and the nucleic acid sequence. We predict that the sensitivity of unwinding processivity to external force is a universal feature that should be observed in all helicases. Our prediction is illustrated using T7 and NS3 helicases as case studies. Interestingly, the increase in unwinding processivity with force depends on whether the helicase forces base pair opening by direct interaction or if such a disruption occurs spontaneously due to thermal fluctuations.
Based on the theoretical results, we propose that proteins like single-strand binding proteins associated with helicases in the replisome, may have co-evolved with helicases to increase the unwinding processivity even if the velocity remains unaffected. \end{abstract}
\pacs{PACS}
\maketitle

\section{Introduction}

Helicases are molecular motors that  can translocate along single-stranded (ss) nucleic acids (NAs)
and can also  unwind double stranded (ds)
NAs when confronted by a single strand-double strand junction \cite{lohman1992,lohman_mechanisms_1996,DelagoutteQRevBiophys02,DelagoutteQRevBiophys03}. These remarkable motors, found in all organisms, are involved in a variety of cellular functions ranging from DNA and RNA replication, remodeling of chromatin, translation and other aspects of nucleic acid metabolism \cite{bianco_2000, lohman_non2008,pyle_translocation_2008, rocak_dead-box_2004, bustamante_revisiting_2011,DongJBiolChem95,VelankarCell97,MariansStructure00,PangEMBO02,VenkatesanJBiolChem82}, and are classified into six families based on their sequence \cite{GorbalenyaCOSB93,IyerJStructBiol04}. Various ensemble assays have been developed \cite{wong_allosteric_1992, ali_kinetic_1997,  levin_helicase_2002, lucius_general_2003, jeong_dna-unwinding_2004} to decipher how they step on single strands of nucleic acids, how they interact with single strand--double strand junctions, how often they dissociate from their track. Single molecule manipulation techniques, such as laser optical tweezers
(LOT) and Magnetic Tweezers (MT), have also been used to  probe the kinetics of stepping and nucleic acid unwinding at the level of individual helicase molecules \cite{PatelARB00,PerkinsBiophysJ04,DessingesPNAS04,LionnetNucAcidsRes06}. These experiments are particularly suited to obtain quantitatve measurements of helicase processivity and unwinding velocity under force. From such measurements, one can obtain insights into the mechanisms of helicase function, and how various helicases differ from each other. 

In a number of single molecule experiments, a variety of interesting observations have been made on the response of helicase velocity and processivity to force. Processivity is a measure of the average number of base-pairs unwound by the helicase in a single binding event (see below for a more detailed discussion of various other definitions).  Dessinges  \textit{et al}.\ \cite{DessingesPNAS04}, who used MT to follow the unwinding of dsDNA by the UvrD helicase, showed that the unwinding velocity  depends only weakly  over a broad range (3-35 pN) of applied forces on the ds termini of the DNA.  However, the unwinding processivity was much larger (265 bp) in their experiment compared to previous results (45 bp) from zero-force ensemble assays \cite{ali_kinetic_1997}, suggesting that \textbf{force possibly plays a part in enhancing the processivity of UvrD.  It should be kept in mind that the stoichiometry of helicase attachment to DNA was not explicitly determined in  \cite{DessingesPNAS04}. This is particularly relevant because  enhanced processivity is achieved by UvrD dimers \cite{maluf_self2003,lee_direct_2013}}.  In sharp contrast, Johnson \textit{et al}.\ \cite{johnson_single-molecule_2007} discovered using LOT experiments, that both the unwinding velocity and processivity of T7 helicase are highly
tension dependent.\ \ Indeed, the unwinding rate increased by an
order of magnitude when the tensile force applied to destabilize the ss-dsDNA junction was increased
from 5 to 11 pN. Finally, Lionnet \textit{et al}.\ \cite{lionnet_real-time_2007}
examined the unwinding kinetics of the gp41 helicase.  
Like the T7 helicase, the unwinding
velocity depends strongly on the value of the tension destabilizing
the ss-dsDNA junction.  From these experiments, we surmise that while different helicases exhibit very different unwinding velocity responses to force, 
the processivity always increases appreciably with force. Here, we provide a theory that explains the differing dependencies of unwinding velocity and processivity on force. 

Some aspects of the varied responses in helicase velocities have been qualitatively justified in a  previous work \cite{manosas_active_2010}, based upon an insightful  theoretical model of helicases originally proposed by Betterton and J\"ulicher (BJ) \cite{betterton_motor_2003,betterton_opening_2005,betterton_velocity_2005}.  The model quantifies the crucial ideas of ``active'' and ``passive'' unwinding by a helicase--a classification that has been the basis for understanding helicase mechanisms for a number of years \cite{lohman1992,lohman_mechanisms_1996, amaratunga_escherichia_1993, von_hippel_general_2001, jeong_dna-unwinding_2004, cheng_ns3_2007, lionnet_real-time_2007, johnson_single-molecule_2007, qi_sequence-dependent_2013, byrd_dda_2012}. A passive helicase utilizes the thermal breathing of the single strand-double strand junction of a nucleic acid, to opportunistically step in front. Since the double strand (ds) closes on itself on average, the helicase frequently faces a barrier and hence moves ahead relatively slowly (as compared to its unimpeded single strand velocity). \textbf{An active helicase on the other hand, promotes dsDNA melting by lowering the stability of base-pairs at the junction, either by utilizing free energy from ATP hydrolysis or the binding free energy of the protein at the junction.} The resulting increase in unwinding velocity of an active helicase depends on the extent of the destabilization of the ds junction due to interaction with the helicase. This suggests that the helicase can be active to various degrees. 

An ``optimally active'' helicase would unwind at a maximum velocity that is close to or equals  the single strand velocity, $V_{\text{SS}}$ \cite{betterton_opening_2005}. For helicases with negligible back stepping rates, it was argued in \cite{manosas_active_2010} that when a helicase is passive, an external force can assist in the opening of the double-strand junction, thus increasing the velocity of the helicase. On the other hand, since the unwinding velocity of an optimally active helicase is $\sim V_{\text{SS}}$,  an external force should not increase the unwinding velocity significantly. These physically motivated arguments suggest that the velocity of a passive helicase should increase appreciably with force while the velocity of an optimally active helicase will be similar to $V_{\text{SS}}$.
Manosas \textit{\textit{et al}} \cite{manosas_active_2010} did not investigate the processivities of helicases, and hence their work provided only an incomplete understanding of the nature of helicase motion. By generalizing the BJ model \cite{betterton_motor_2003,betterton_opening_2005,betterton_velocity_2005} to include force dependence for both velocities and processivities, we show that even though the velocity can vary with external force in a helicase--dependent manner, the unwinding processivity always increases significantly with force. This is a remarkable result, as it predicts a universal behavior --- unlike the velocity, the processivity will increase rapidly with external force for all helicases regardless of their architecture. In addition, our work highlights another surprising fiding --- the more active a helicase is (the stronger its interaction with the single strand-double strand junction), the less processive is its motion. \textbf{The effect of increasing the external force at a given interaction strength is therefore, opposite to the effect of increasing the interaction strength at a given external force.} Finally, the sequence dependent behavior of the unwinding velocity and processivity show complex behavior depending on the  percentage of GC content. We predict that details of the energy landscape of base pair opening, GC content, and the extent to which the helicase is active determine the unwinding velocity and processivity. 

\section{Materials and Methods}

\begin{figure}[h]
\begin{center}
\includegraphics[width=\columnwidth]{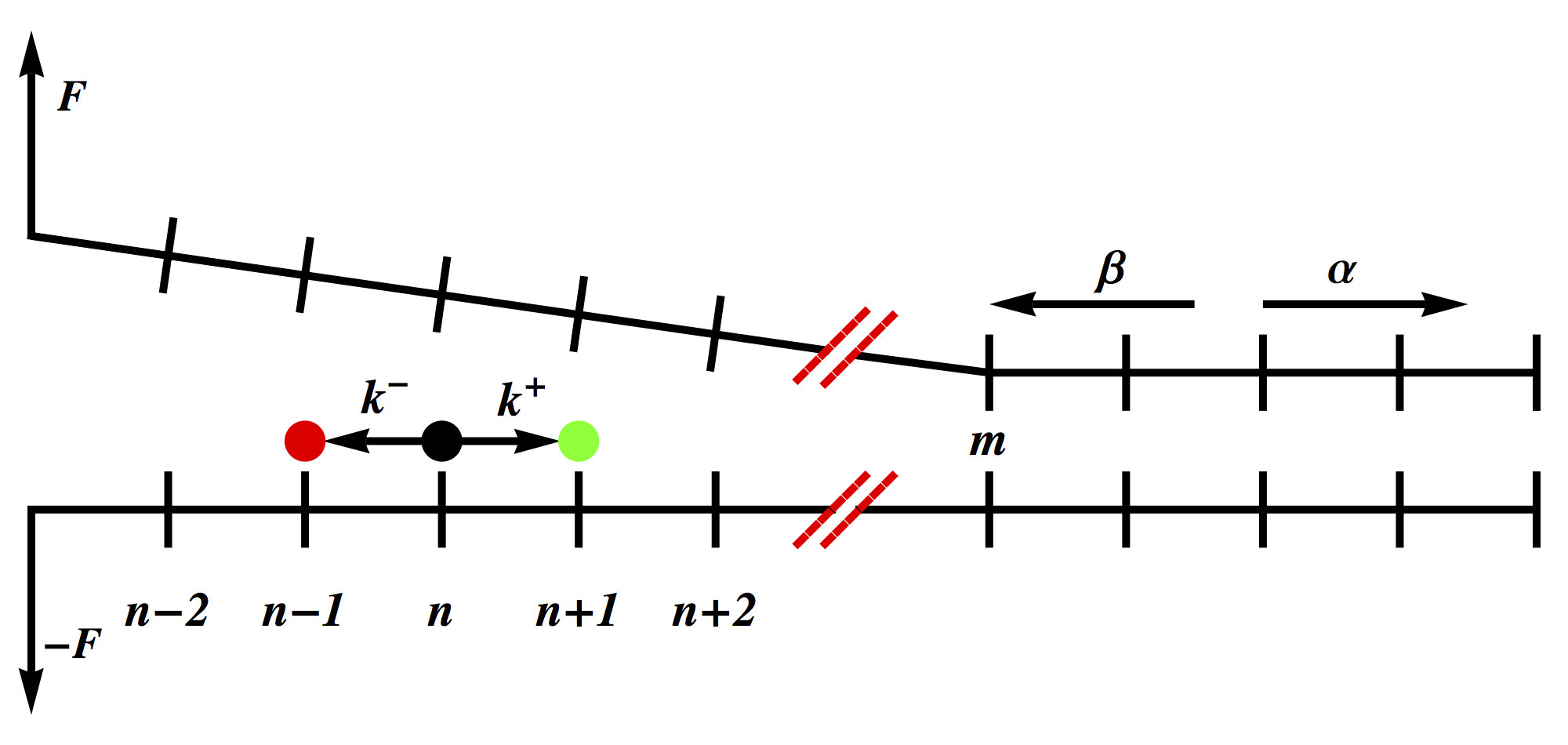}
\end{center}
\caption{A schematic illustration of the extension of the Betterton and
J\"ulicher model for helicases. \ \ The position of the
helicase (black circle) on an underlying 1D lattice representing
the nucleic acid substrate is denoted by the variable $n$, while
the variable $m$ refers to the location of the ss-dsNA junction. 
\ \ At infinite separation between the helicase and ss-ds NA junction, $n\rightarrow n+1$ transitions occur at rate $k^{+}$,
while $n\rightarrow n-1$ transitions occur at rate $k^{-}$.\ \ Similarly, $m\rightarrow
m+1$ transitions occur at rate $\alpha $, and $m\rightarrow m-1$ transitions
occur at rate $\beta $.\ \ The tension $F$ is applied to the
ends of the nucleic acid.}
\label{XRef-FigureCaption-42611052}
\end{figure}

As illustrated schematically in Fig.\ \ref{XRef-FigureCaption-42611052}, in the BJ model the nucleic acid (NA) is represented by a one-dimensional lattice with $n$
denoting the lattice position of the helicase, while $m$ specifies
the position of the ss-ds junction. {\bf Models similar in spirit, but differing significantly in formulation and application, have also been used to analyze motility of kinesin \cite{kolomeisky_periodic_2000}}. We extended the BJ model to include a constant external force or tension ($F$), which is applied to the complementary termini of the NA, in order to investigate the dependence of velocity and processivity on load. \ \ In accord with the
single molecule optical tweezers experiments \ \cite{DumontNature06, johnson_single-molecule_2007}, tension is applied in a manner that increases
the opening rate of the junction and decreases the closing rate while maintaining detailed balance.

If the helicase and ss-ds junction are in proximity, there is an effective interaction. The passive unwinding mechanism is realized by a hard-wall coupling
potential:
\begin{equation}
\label{XRef-Equation-416152417}
U(j) =
\begin{cases}
\infty &\left( j\leq 0\right)\\
0 &\left( j>0\right),
\end{cases}
\end{equation}

\noindent where $j \equiv m-n$ (Fig.~\ref{XRef-FigureCaption-42611052}).  Similarly, we represent an active unwinding mechanism by a coupling potential
consisting of both a step-function and a hard-wall:
\begin{equation}
\label{XRef-Equation-427105745}
U(j) =
\begin{cases}
\infty &\left( j\leq -1\right)\\
U_{0} & \left( j=0 \right)\\ 
0 &\left( j>0\right),
\end{cases}
\end{equation}
where the energy $U_0$ is in units of $k_BT$.

\emph{Unwinding velocity of a helicase.}
The mean velocity ($V$) of the helicase/junction complex is proportional
to a sum over all $j$ of a product of the probability of
being at separation $j$ and the net rate
at which the centroid coordinate ($l=m+n$, Fig.\ \ref{XRef-FigureCaption-42611052}) increases:
\begin{equation}
V=\frac{1}{2}\sum \limits_{j}\left( k_{j}^{+}+\alpha _{j}-k_{j}^{-}-\beta
_{j}\right) P_{j},%
\label{XRef-Equation-416152536}
\end{equation}
\noindent where $\alpha _{j}$ is the rate constant associated with junction
opening when the helicase and junction are at separation $j$.  Similarly,
$\beta _{j},k_{j}^{+}$, and $k_{j}^{-}$, are the $j$-dependent rate
constants associated with junction closing, helicase forward stepping,
and helicase backward stepping respectively.

Since experiments are performed at a constant temperature and $F$, opening
and closing of the ss-ds junction can only be attributed to thermal
fluctuations or the applied tension.\ \ Thus, the ratio of the rate
at which the junction opens ($\alpha_{j-1}$) to that at which it closes ($\beta_{j}$) satisfies:
\begin{eqnarray}
\frac{\alpha _{j-1}}{\beta _{j}} &=&\frac{\alpha }{\beta } e ^{\left[
U( j-1) -U( j) \right] } e ^{\Delta G_F} \nonumber \\
\frac{\alpha}{\beta} &=& e^{-\Delta G},
\label{XRef-Equation-416145310}
\end{eqnarray}

\noindent where $\alpha$ ($\beta$)  is the $j\rightarrow \infty $ junction opening
(closing) rate, $\Delta G$ is the stability of the junction base pair in the absence of either force or helicase, $U( j)$ is the value of the coupling potential when the helicase
and junction are at separation $j$, and $\Delta G_F$ is the destabilization in the free energy of the basepair at the junction, caused by the applied force $F$. For simplicity, we choose $\Delta G_F=F \Delta x$ (see further discussions in the Discussion section), where $\Delta x$ is roughly twice the length of ssDNA separating two bases.
When ATP hydrolysis is tightly coupled to  helicase transitions,
the hopping rates approximately satisfy a relation akin to detailed balance \cite{betterton_motor_2003}: 
\begin{equation}
\frac{k_{j}^{+}}{k_{j-1}^{-}}\approx \frac{k^{+}}{k^{-}} e ^{-\left[
U( j-1) -U( j) \right] },%
\label{XRef-Equation-41615721}
\end{equation}
\noindent where $k^{+}=\operatorname*{\lim }\limits_{j\,\rightarrow
\:\infty }k_{j}^{+}$ and $k^{-}=\operatorname*{\lim }\limits_{j\,\rightarrow
\:\infty }k_{j-1}^{-}$. Note that when the helicase traverses on a single strand, 
there is no single strand--double strand junction confronting the helicase, and hence 
the hopping rates that describe its motion are $k^+$ and $k^-$. Thus, the 
single strand velocity is $V_{\text{SS}}=k^+-k^-$. \textbf{For a number of helicases like T7, T4, NS3, UvrD which translocate directionally on ss nucleic acids, the forward rate $k^+$ can be assumed to be much larger than $k^-$, near the physiological ATP concentrations ($\sim 1$ mM, close to saturating conditions for most helicases) that are usually used in experiments \cite{jeong_dna-unwinding_2004,lionnet_real-time_2007,johnson_single-molecule_2007,cheng_ns3_2007,manosas_active_2010}. We therefore work under this approximation throughout this article. While unwinding the double-strand however, the forward rate decreases and the backward rate increases (as evident from Eq.~\ref{XRef-Equation-41615721} and the equations for the individual rates below), making back-stepping much more likely. A high probability of backstepping was indeed observed in a recent experiment on the XPD helicase \cite{qi_sequence-dependent_2013}.} 
Finally, individual rates are affected by the coupling potential and the applied tension as follows:
$k_{j}^{+}=k^{+} e ^{-f[ U( j-1) -U( j) ] }$, $k_{j-1}^{-}=k^{-} e ^{-\left( f-1\right) [ U( j-1) -U( j) ] }$, $\beta _{j}=\beta  e ^{-f[ U( j-1) -U( j) +F \Delta x] }$, and $\alpha _{j-1}=\alpha  e ^{-\left( f-1\right) [ U( j-1) -U( j) +F\Delta x] }$, where $f \, (0<f<1)$ is the location (in units of lattice spacing) of the transition state separating the closed and open states.
Note that the applied tension $F$, affects only the nucleic
acid breathing rates ($\alpha _{j-1}$ and $\beta _{j}$), but not  the hopping rates.

The rates of a passive helicase are independent of $j$ for  $j>1$, and at $j=1$ we have $k_{1}^{+}=\beta _{1}=0$.\ \ The force-velocity relation for a `hard-wall' (passive) helicase, calculated using Eq.\ \ref{XRef-Equation-416152536}, is:
\begin{equation}
V_{\mathrm{HW}\text{}} =\frac{\alpha ^{\prime }k^{+}-\beta ^{\prime
}k^{-}}{\beta ^{\prime }+k^{+}},%
\label{XRef-Equation-416153226}
\end{equation}

\noindent where $V_{\mathrm{HW}\text{}} \equiv V_{\mathrm{HW}\text{}}(F, f)$, $\alpha ^{\prime }\equiv \alpha  e ^{-(f-1)F \Delta
x}$ and $\beta ^{\prime }\equiv \beta  e ^{-f F \Delta x}$. For convenience, while referring to $V_{\mathrm{HW}\text{}}$ in the rest of the text, we explicitly show the functional dependence of only the parameter pertinent to the particular discussion. Note
that for $F=0$,\ \ Eq.\ (\ref{XRef-Equation-416153226}) coincides
with Eq.\ (22) in reference \cite{betterton_opening_2005}.\ \ By making the
approximation $k^{-}\approx 0$, we obtain $V_{\mathrm{HW}} \approx \frac{\alpha ^{\prime }}{\beta ^{\prime
}}\left( \frac{k^{+}}{1+k^{+}/\beta ^{\prime }}\right)$ and 
$\frac{V_{\mathrm{HW}}( F) }{V_{\mathrm{HW}}( F=0) }\approx \frac{\alpha
^{\prime }/\alpha }{\beta ^{\prime }/\beta }\left( \frac{1+k^{+}/\beta}{1+k^{+}/\beta ^{\prime }}\right)$.
Finally, making the approximation $\beta ^{\prime
}\gg k^{+}$, we conclude that:
\begin{equation}
\frac{V_{\mathrm{HW}}( F) }{V_{\mathrm{HW}}( F=0) }\approx  e ^{F
\Delta x}.%
\label{XRef-Equation-4161727}
\end{equation}

\noindent (The approximation $\beta ^{\prime }\gg k^{+}$ is
typically valid because $\beta \simeq {10}^{5}$-${10}^{8} s^{-1}$,
while $k^{+ }\simeq {1}$-${10}^{3} s^{-1}$).\ \ A passive helicase must wait for the junction to open to step forward and prevent the newly separated base pair from reannealing.  From Eq.~\ref{XRef-Equation-4161727} it follows that the application of force exponentially increases the probability that the junction is open relative to the probability that it is closed, resulting in an exponential increase in the unwinding velocity relative to the $F=0$ value. 

For a single step active helicase (modeled with Eq.\ \ref{XRef-Equation-427105745}) the rates are independent of $j$ for all
$j>1$.
A straightforward calculation of Eq.~\ref{XRef-Equation-416152536} leads to
\begin{equation}
\frac{V_{1} }{V_{\mathrm{HW}} }= \frac{c^{\prime }+\left(
1-c^{\prime }\right)  e ^{-f U_{0}}}{c^{\prime }+\left( 1-c^{\prime
}\right)  e ^{-U_{0}}},%
\label{XRef-Equation-419141842}
\end{equation}

\noindent where $V_1 \equiv V_1(F, f, U_0)$, $c^{\prime } =(\alpha  e ^{-(f-1)F \Delta x}+k^{-})/(k^{+}+\beta
e ^{-f F \Delta x})$ and the subscripts $1$ and $\mathrm{HW}$ denote
the step (active) and hard-wall (passive) coupling potentials. When the step potential goes to zero ($U_0=0$), the unwinding velocity is just equal to the hard wall (passive) helicase velocity as can be seen from Eq.~\ref{XRef-Equation-419141842}.  As with $V_{\mathrm{HW}}$, while referring to $V_1$ in the rest of the text, we show only the functional dependence of the parameters pertinent to the particular discussion. Finally, using the results of Eqs.\ \ref{XRef-Equation-4161727} and \ref{XRef-Equation-419141842} we find:
$\frac{V_{1}( F) }{V_{1}( F=0) }=\left( \frac{c^{\prime
}+\left( 1-c^{\prime }\right)  e ^{-f U_{0}}}{c^{\prime }+\left(1-c^{\prime }\right)  e ^{-U_{0}}}\right)  e ^{F \Delta x}( \frac{c+\left(1-c\right)  
e ^{-U_{0}}}{c+\left( 1-c\right)  e ^{-f U_{0}}})$, 
where $c=c^{\prime }( F=0) =(\alpha +k^{-})/(k^{+}+\beta
)$.

\emph{Helicase processivity.}  
The processivity of a helicase has been defined in various ways \cite{betterton_velocity_2005}:
(1)  the mean attachment time of the helicase $\langle \tau \rangle
$,
(2)  the average number of base pairs unwound in a single binding
event $\langle \delta m\rangle$ (in other words, the average number of base pairs by which the junction moves ahead before the helicase detaches), and
(3)  the average number of base pairs translocated
before the helicase detaches $\langle \delta n\rangle$; $\langle \delta m\rangle$ and 
$\langle \delta n\rangle$ (termed unwinding and translocation processivities respectively) in principle  could 
be very different. For instance, if the helicase binds very far from the junction ($j_0 \gg 1$), the double strand will close 
rapidly before the helicase can translocate by a significant amount, thus making the unwinding processivity ($\langle \delta m \rangle$) negative.
However, if the helicase binds very close to the junction ($j_0 \sim 1$), $\langle \delta m\rangle$  and $\langle \delta n\rangle$ are 
almost identical \cite{betterton_velocity_2005}. The double strand would always have a 
larger closing rate as compared to the opening rate, thus making the strands close on average. This would hold even 
in the presence of external forces, as long as they are less than the force needed to unzip the double strand. In such physically 
relevant situations, after a very brief transition period the helicase is likely to be very close to the junction, and hence the relevant initial condition can always be taken as $j_0=1$. 

In the single molecule experiments measuring unwinding velocity and processivity, typically a DNA hairpin serves as 
a model double strand. The upstream single strand overhangs allow the helicase to load \cite{manosas_active_2010,cheng_ns3_2007, lionnet_real-time_2007,DumontNature06}. At forces less than the critical force required to unzip the hairpin in the absence of a helicase, the arrival of a helicase at the junction and subsequent unwinding causes the end-to-end distance to increase. As a result, the presence of the helicase on the hairpin can only be discerned by observing the sudden change in the end-to-end distance, which happens when $j_0 \sim 1$. In this work therefore, we work with the initial condition  $j_0=1$, and hence the unwinding processivity $\langle \delta m\rangle$ and translocation processivity $\langle \delta n\rangle$ are almost identical. 

To model a helicase with finite processivity, we incorporate an
unbinding rate $\gamma _{j}$ which depends implicitly on the separation
$j$ through the relation $\gamma _{j}=\gamma  e ^{U( j) }$ \cite{betterton_velocity_2005}.\ \ We
assume that $U( j) \rightarrow 0$ as $j\rightarrow \infty $. \textbf{It is physically reasonable that the unbinding rate should increase as $U(j)$ increases, because the repulsion between the helicase and the nucleic acid would cause the former to dissociate. This has indeed been observed in experiments --- the dissociation rate of T7 from dsDNA is roughly 100 times larger \cite{jeong_dna-unwinding_2004} than that from ssDNA \cite{kim_t7_2002}, at the same dTTP concentration of 2mM. The exponential dependence of $\gamma _{j}$ on $U( j)$ is further justified by noting that it is built on the celebrated Bell model, which has been used successfully to describe unbinding of cell-adhesion moleclues and other complexes.} Since the 
velocity of unwinding is unaffected by introduction of the unbinding rate $\gamma_j$, the analytical results derived above for the unwinding velocity hold good even for finitely processive 
helicases. For the physically relevant initial condition $j_0 \sim1$ (explained above), the velocity of active unwinding is given by: 
\begin{equation}
V_1 \approx \langle \delta m\rangle/\langle \tau\rangle \approx \langle \delta n\rangle/\langle \tau\rangle.
\label{vel}
\end{equation}
For the initial condition $j_0 \sim 1$, the two expressions for unwinding velocity given by Eq.~\ref{XRef-Equation-419141842} and Eq.~\ref{vel} are equivalent. Betterton and J\"ulicher \cite{betterton_velocity_2005} derived the following expressions for the three measures of processivity:
\begin{eqnarray}
\langle \tau \rangle &=& \sum \limits_{j} R_j \nonumber \\
\langle \delta m \rangle &=& \frac{\langle \delta l \rangle + \langle \delta j \rangle}{2} \nonumber \\
\langle \delta n \rangle &=& \frac{\langle \delta l \rangle-\langle \delta j \rangle}{2}
\label{proc}
\end{eqnarray}
where the parameters $\{R_{j}\}$ are obtained by solving the following infinite set of second order recurrence relations: 
\begin{equation}
-\delta _{j j_{0}}=-\left( k_{j}^{+}+k_{j}^{-}+\alpha _{j}+\beta _{j}+\gamma _{j}\right)
R_{j}+\left( \alpha _{j-1}+k_{j-1}^{-}\right) R_{j-1}+\left( \beta
_{j+1}+k_{j+1}^{+}\right) R_{j+1},%
\label{XRef-Equation-419155226}
\end{equation}
 $j_{0}$  is the value of $j$ at time $t=0$, $\delta _{j j_{0}}$ is the Kronecker delta, $\langle \delta j \rangle=\sum_j (j-j_0) \gamma_j R_j$, and $\langle \delta l \rangle$  is given by the expression:
$\langle \delta l \rangle=b \left ( 1+a-y_- -\frac{a-b}{y_+} \right )^{-1} \left [ \frac{y_-}{(1-y_-)^2}-\frac{y_+}{(1-y_+)^2} \right ]$, where $y_{\pm}$ are the roots of the equation $y^2-(1+a)y+(a-b)=0$, $a=(1+p)/q$, $b=1/q$ and $p=\sum_j (\alpha_j+k_j^+)R_j$, $q=\sum_j(\beta_j+k_j^-)R_j$. BJ derived Eq.~\ref{XRef-Equation-419155226} by taking the Laplace Transform of the time evolution equation of $P(j,l,t)$, the joint probability density of finding the helicase-junction system in state $(j,l)$ at time $t$. $P(j,l,t)$ obeys a Master equation accounting for the movement of the helicase in the forward and backward direction as well as detachment from the NA.

When the helicase is passive, $U(j)=0$ for $j>0$ (Eq.~\ref{XRef-Equation-416152417}) and hence the dissociation rate $\gamma_j=\gamma$ for all $j$. As a result, the mean attachment time for a passive helicase $\langle \tau \rangle_\text{HW}$ is given by
\begin{equation}
\langle \tau \rangle_{\text{HW}}=\frac{1}{\gamma} \, ,
\label{tauhw}
\end{equation}  
for all values of the external force $F$. 

\section{Results}
\begin{figure}[h]
\begin{center}
\includegraphics[width=\columnwidth]{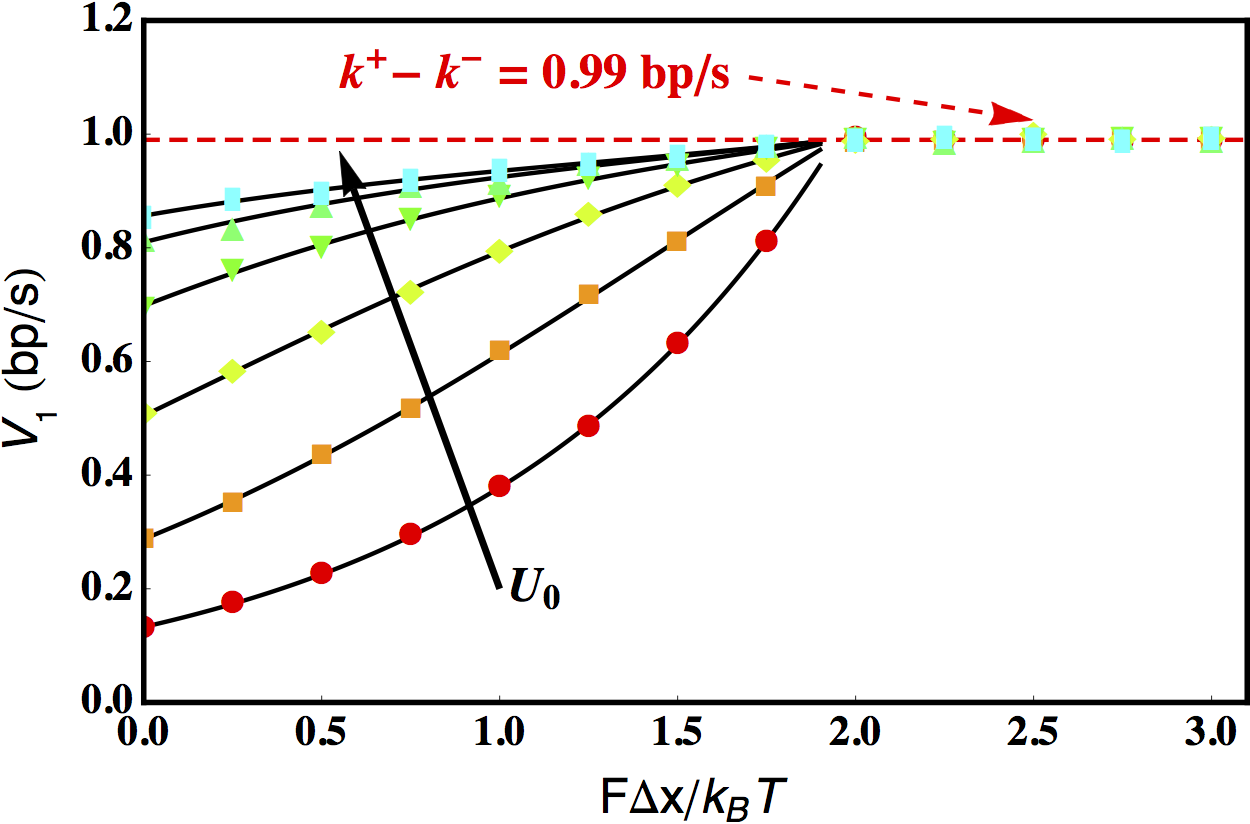}
\end{center}
\caption{Unwinding velocity $V_1(F, U_0)$ as a function of the tension ($F \Delta x/k_{B}T$), for various coupling potentials $U_{0}/k_{B}T=
0$ (red circles), $1$ (orange squares), $2$ (yellow diamonds), $3$ (green
downward triangles), $4$ (green upward triangles), $5$ (cyan rectangles).\ \ The
solid black lines are numerical results obtained using Eq.~\ref{vel} and Eq.~\ref{XRef-Equation-419155226}. Each symbol
represents an average of 1000 independent Kinetic Monte Carlo simulations.\ \ For forces exceeding $F \Delta x/k_{B}T=1.9 $, the duplex
melts and it is no longer possible
to numerically solve the system given by Eq.\ \ref{XRef-Equation-419155226}, but the robustness of our simulation algorithm
allows us to explore this regime confirming that the duplex
has melted.\ \ The parameters used were $f=0.01$, $\alpha ={10}^{5}
s^{-1}$, $\beta =7\times {10}^{5} s^{-1}$, $k^{+}=1 \mathrm{bp}/s$,
$k^{-}=0.01 \mathrm{bp}/s$, and $\gamma =0.01 s^{-1}$.}
\label{XRef-Figure-422171440}
\end{figure}

In order to determine the velocity and  processivity, we solve the sparse linear
system given by Eq.\ \ref{XRef-Equation-419155226} for $\{R_{j}\}$. \textbf{All our results have been numerically obtained for the case where the back-stepping rate of the helicase in the absence of dsDNA ($k^-$) is small compared to the forward rate in the absence of dsDNA ($k^+$).} We
use a grid of size $M=10,000$ in $j$ to solve Eq.\ \ref{XRef-Equation-419155226}.\ \ A
larger grid size than that used by BJ ($M=100$ was used in Ref.
\cite{betterton_velocity_2005} ) is necessitated by the external tension
applied to the NA substrate.\ \ From a numerical standpoint, as
$F \Delta x/k_{B}T$ increases, $\{R_{j}\}$ converges to zero increasingly
slowly, thus requiring more and more terms  to
guarantee the convergence of the sums used to calculate $\left\langle \tau \right\rangle$, $\left\langle  \delta  j\right\rangle$,
 and $\left\langle  \delta  l\right\rangle$.
For $F \Delta x/k_{B}T>1.95$, a grid
of size $M={10}^{6}$ in $j$ proves to be insufficient to solve
Eq.\ (\ref{XRef-Equation-419155226}).\ \ It is not possible to explore
forces greater than 1.95 $k_{B}T/\Delta x$ because such large forces result in the melting of the duplex. This is because our choice of $\alpha=10^5 s^{-1}$ and $\beta=7 \times 10^5 s^{-1}$ correspond to $\Delta G=1.95 k_BT$. \ \ At these forces, $\langle \delta m\rangle $ is very large
and $\langle \delta n\rangle \approx \langle \tau \rangle (k^{+}-k^{-})$.\ \ These
results are confirmed in Fig.\ \ref{XRef-Figure-422171440}, which
shows (in addition to the numerical results) simulation results
at these larger forces (see below for additional information).

\emph{Unwinding velocity}: To understand how the unwinding velocity depends on the model parameters, we plot the $F$-dependent unwinding velocity $V_1(F, U_0)$  for several values of $U_0$ (Fig.\ \ref{XRef-Figure-422171440}). This figure reveals  several interesting features.
(1) Fig.\ \ref{XRef-Figure-422171440} shows that for a small value of $f=0.01 \ll1$  ($f$ is the transition state location along the reaction coordinate separating the closed and open states of a base pair), if the helicase is passive $(U_0=0)$ or weakly active ($U_0<\Delta G$), the unwinding velocity is highly sensitive to the external force. As $U_0$ increases, making the helicase increasingly more active, the unwinding velocity is less sensitive to increase in force. In the parameter range $f<<1$ and $U_0/k_BT>>1$, the helicase is ``optimally active''. (2) The parameters used in Fig.\ \ref{XRef-Figure-422171440} correspond to a single-strand translocation velocity $V_{\text{SS}}$=0.99 bp/s. In the optimally active regime, the unwinding velocity is close to $V_{\text{SS}}$ for a large range of forces. This observation along with point (1) qualitatively describe the velocity behavior of a number of helicases 
\cite{manosas_active_2010}. For example, UvrD is optimally active while T7 is passive or at best, only weakly active. (3)  When $U_{0}=0$, the results revert
to those expected for a hard wall coupling potential.
Thus, as illustrated in Fig.\ \ref{XRef-FigureCaption-42611052}, the effect
of tensile force at $U_{0}=0$ is simply to increase the mean velocity
exponentially relative to the zero force value (Eq.~\ref{XRef-Equation-4161727}).

\begin{figure}[h]
\begin{center}
\includegraphics[width=\columnwidth]{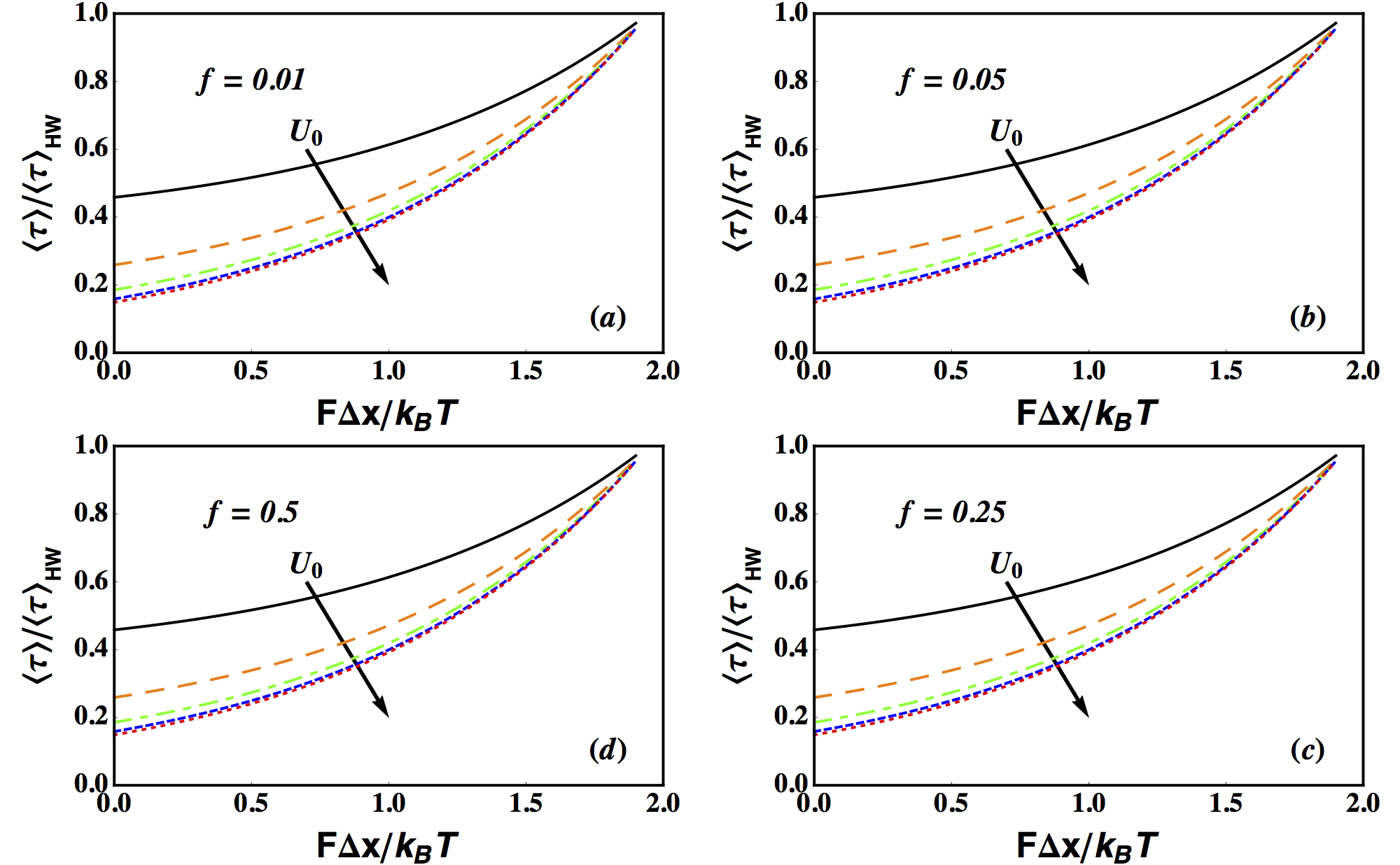}
\end{center}
\caption{Ratio of the mean attachment time of an active to a passive helicase
$\langle \tau \rangle /{\langle
\tau \rangle }_{\mathrm{HW}}$ as a function of $F \Delta x/k_{B}T$, for
$U_{0}/k_{B}T=1$ (black solid line), $2$ (orange dashed line), $3$ (green
short and long dashed line), $4$ (blue short dashed line), $5$ (red
dotted line).\ \ $\langle \tau \rangle /{\langle \tau \rangle }_{\mathrm{HW}}$ increases
with $F \Delta x/k_{B}T$ and decreases with increasing
$U_{0}/k_{B}T$.\ \ Interestingly, the mean attachment
time is unaffected by $f$ (all curves with the same value of $U_{0}/k_{B}T$ can be superimposed).\ \
Parameters used to solve Eq.\ \ref{XRef-Equation-419155226}
were $\alpha ={10}^{5} s^{-1}$, $\beta =7\times {10}^{5} s^{-1}$,
$k^{+}=1 \mathrm{bp}/s$, $k^{-}=0.01 \mathrm{bp}/s$, $\gamma =0.01
s^{-1}$, $j_{0}=1$, and $M=10^4$.}
\label{XRef-FigureCaption-422164254}
\end{figure}

\begin{figure}
\begin{center}
\includegraphics[width=\columnwidth]{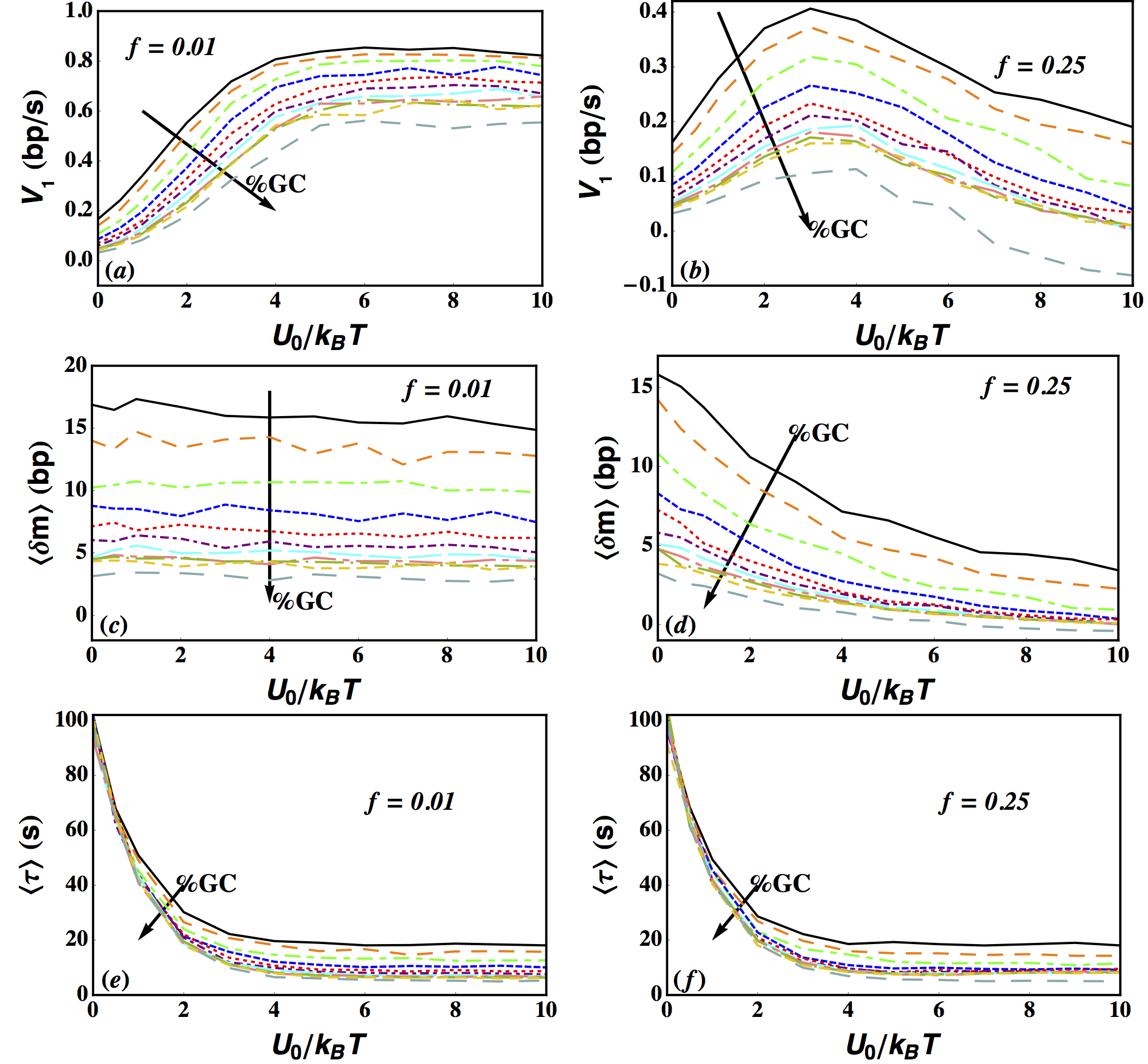}
\end{center}
\caption{The dependence of $\langle \tau \rangle $, $\langle \delta m\rangle
$, and $\langle \delta m\rangle /\langle \tau \rangle $ on the
step-height ($U_{0}/k_{B}T$) for varying amounts of GC content for the sequence given in the text.\ \ Simulations were performed at $U_{0}/k_{B}T=0,
0.5,1, 2, 3, 4, 5,6,7,8,9,$ and $10$.\ \ Each data point corresponds
to an average over 1000 independent Kinetic Monte Carlo simulations.\ \ All quantities decrease with
increasing \%GC.\ \ Interestingly, the mean attachment time $\langle
\tau \rangle $ is again very insensitive to $f$.\ \ $\langle \delta
m\rangle $ is, however, sensitive to $f$, leading to disparate
behaviors for {\textlangle}$\delta m\rangle /\langle \tau \rangle
$.\ \ When $f=0.25$, $\langle \delta m\rangle /\langle \tau \rangle
$ shows a very distinct maximum when \%GC=0.0 but a very weak maximum
when \%GC=1.0.\ \ When $f=0.01$, $\langle \delta m\rangle /\langle
\tau \rangle$ shows saturating behavior with increasing $U_{0}$.\ \ Thus,
sequence plays a crucial role in determining the kinetics of
unwinding.\ \ Parameters used in the simulations were $\alpha ={10}^{5}
s^{-1}$, $\beta =7\times {10}^{5} s^{-1}$, $k^{+}=1 \mathrm{bp}/s$,
$k^{-}=0.01 \mathrm{bp}/s$, $\gamma =0.01 s^{-1}$,\ \ $j_{0}=1$.}
\label{XRef-FigureCaption-422164915}
\end{figure}

\emph{Processivity and lifetimes.} For finitely processive helicases, the primary results of numerically solving Eq.~\ref{XRef-Equation-419155226} are plotted in Figs.\
\ref{XRef-FigureCaption-422164254}, \ref{XRef-FigureCaption-422164312}, \ref{up} and 
 \ref{XRef-FigureCaption-422164915}, illustrating several points worthy of note: 
\begin{enumerate}
\item The variations in the average lifetime of the helicase $\langle \tau \rangle$ (normalized by the lifetime of 
the passive helicase) with $F$, $f$ and $U_0$ (Eq.~\ref{proc}) are displayed in  Fig.\ \ref{XRef-FigureCaption-422164254}. Evidently, a passive helicase ($U_0=0$)  has the largest lifetime. As $U_0$ increases, lifetimes monotonically decrease at any given 
force value. This is not surprising since the lifetime is controlled by the detachment rate $\gamma_j$, which increases exponentially  with increase in $U_0$. 
As $F$ increases, the opening rate of the double strand becomes larger, thus increasing the probability of the helicase to find an open adjacent base whenever it steps ahead. This results in fewer occasions where the helicase has to pay the extra $U_0$ energy to plough ahead, resulting in larger lifetimes. At forces close to the double-strand rupture force, the helicase always finds a clear path ahead and hence rarely interacts with the junction, making the lifetime increasingly approach the passive helicase lifetime. Finally,  Fig. \ref{XRef-FigureCaption-422164254} also demonstrates that the mean attachment time of the helicase relative to that of a passive helicase depends on both $U_{0}$ and $F$,
but is insensitive to the parameter $f$.

\begin{figure}[h]
\begin{center}
\includegraphics[width=\columnwidth]{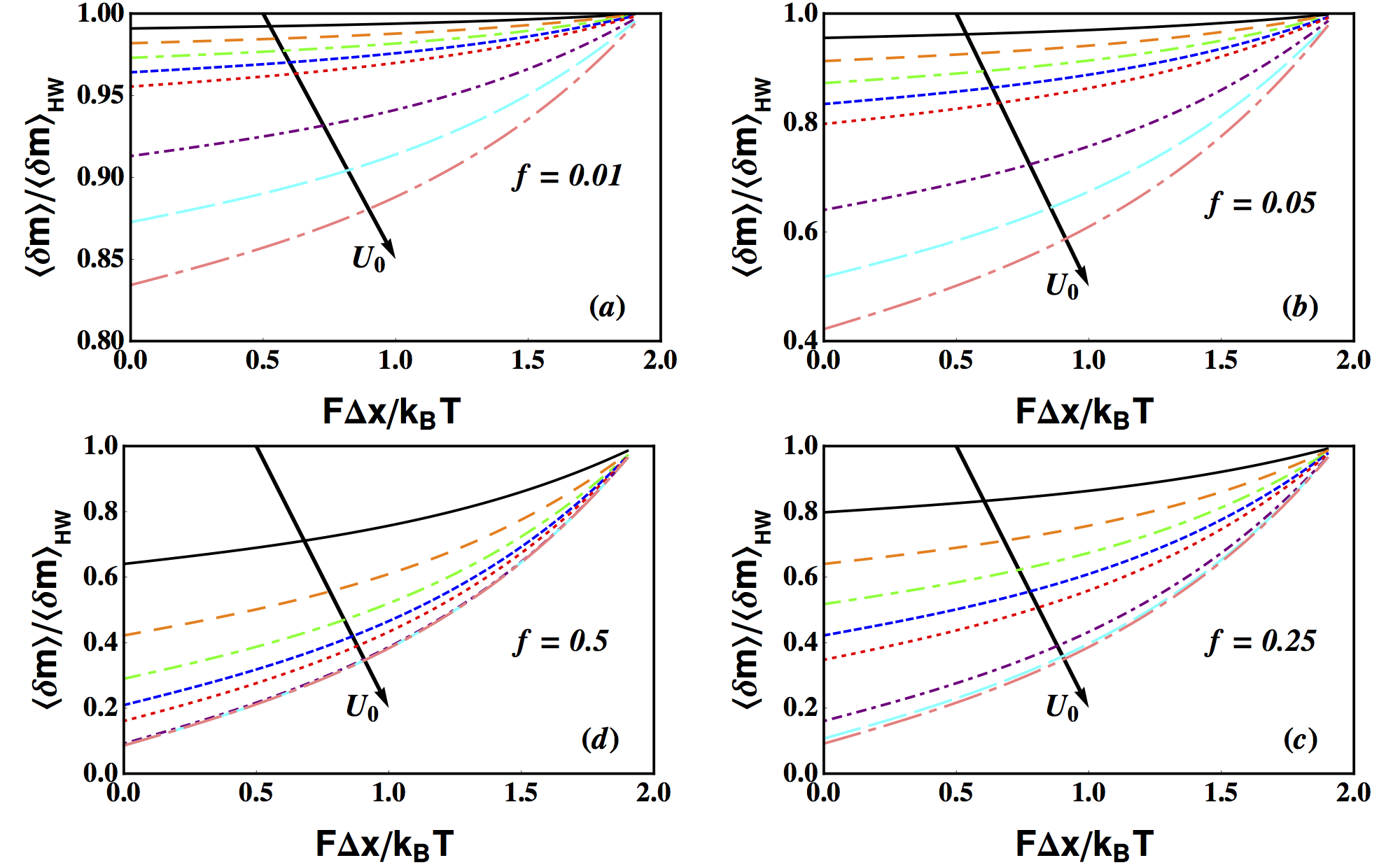}
\end{center}
\caption{Mean unwinding processivity of an active helicase
(relative to that of a passive helicase) as a function of $F \Delta x/k_{B}T$ , for
$U_{0}/k_{B}T=$1, 2, 3, 4, 5, 10, 15, and 20.\ \ For all values of $U_0$, the processivity always
increases with increasing tension destabilizing the ss-dsNA junction
and decreases with increasing step height.\ \ Unlike the mean attachment
time (Fig.\ \ref{XRef-FigureCaption-422164254}), the unwinding
processivity is highly sensitive to the kinetic parameter $f$, further confirming
that the processivity is likely to exert a strong influence over
the kinetics of unwinding. Parameters used to solve Eq.\ \ref{XRef-Equation-419155226}
were $\alpha ={10}^{5} s^{-1}$, $\beta =7\times {10}^{5} s^{-1}$,
$k^{+}=1 \mathrm{bp}/s$, $k^{-}=0.01 \mathrm{bp}/s$, $\gamma =0.01
s^{-1}$,\ \ $j_{0}=1$, and $M=10^4$.}
\label{XRef-FigureCaption-422164312}
\end{figure}

\item  Fig.\ \ref{XRef-FigureCaption-422164312} shows the mean unwinding processivity of an active helicase relative to a passive helicase. Since the translocation processivity is very similar, we do not show a separate figure and all results discussed for $\left\langle  \delta m\right\rangle$, also hold good for $\left\langle  \delta n\right\rangle$. It is evident from the figure that similar to the lifetime (Fig. \ref{XRef-FigureCaption-422164254}), the unwinding processivity is also maximum for a passive helicase, and monotonically decreases as the helicase becomes more active. This interesting result can be physically understood as follows: $\left\langle  \delta m\right\rangle$ depends on the unwinding velocity and mean lifetime as $\left\langle  \delta m\right\rangle= V_1(F,U_0) \left\langle  \tau \right\rangle$. At a given force, as $U_0$ increases, $V_1(F,U_0)$ initially increases (see Fig.\ \ref{XRef-FigureCaption-422164915}a,b) but $\left\langle  \tau \right\rangle$ decreases (see Fig.\ref{XRef-FigureCaption-422164254}, Fig.\ \ref{XRef-FigureCaption-422164915}e,f). The rate at which these two quantities increase/decrease determines the trend for $\left\langle  \delta m\right\rangle$. Our results show that the rate of decrease of $\left\langle  \tau \right\rangle$ is faster than the rate at which $V_1(F,U_0)$ increases. This can be most clearly seen in Fig.\ \ref{XRef-FigureCaption-422164915} (b and f) where the velocity and mean processivities have been plotted as functions of $U_0$ respectively for $f=0.25$. $\left\langle  \tau \right\rangle$ decreases faster than $V_1(F,U_0)$ increases, and hence the overall result is that $\left\langle  \delta m\right\rangle$ decreases as a function of $U_0$ regardless of the GC content. (Fig.\ \ref{XRef-FigureCaption-422164915}d).

\begin{figure}[h]
\begin{center}
\includegraphics[width=\columnwidth]{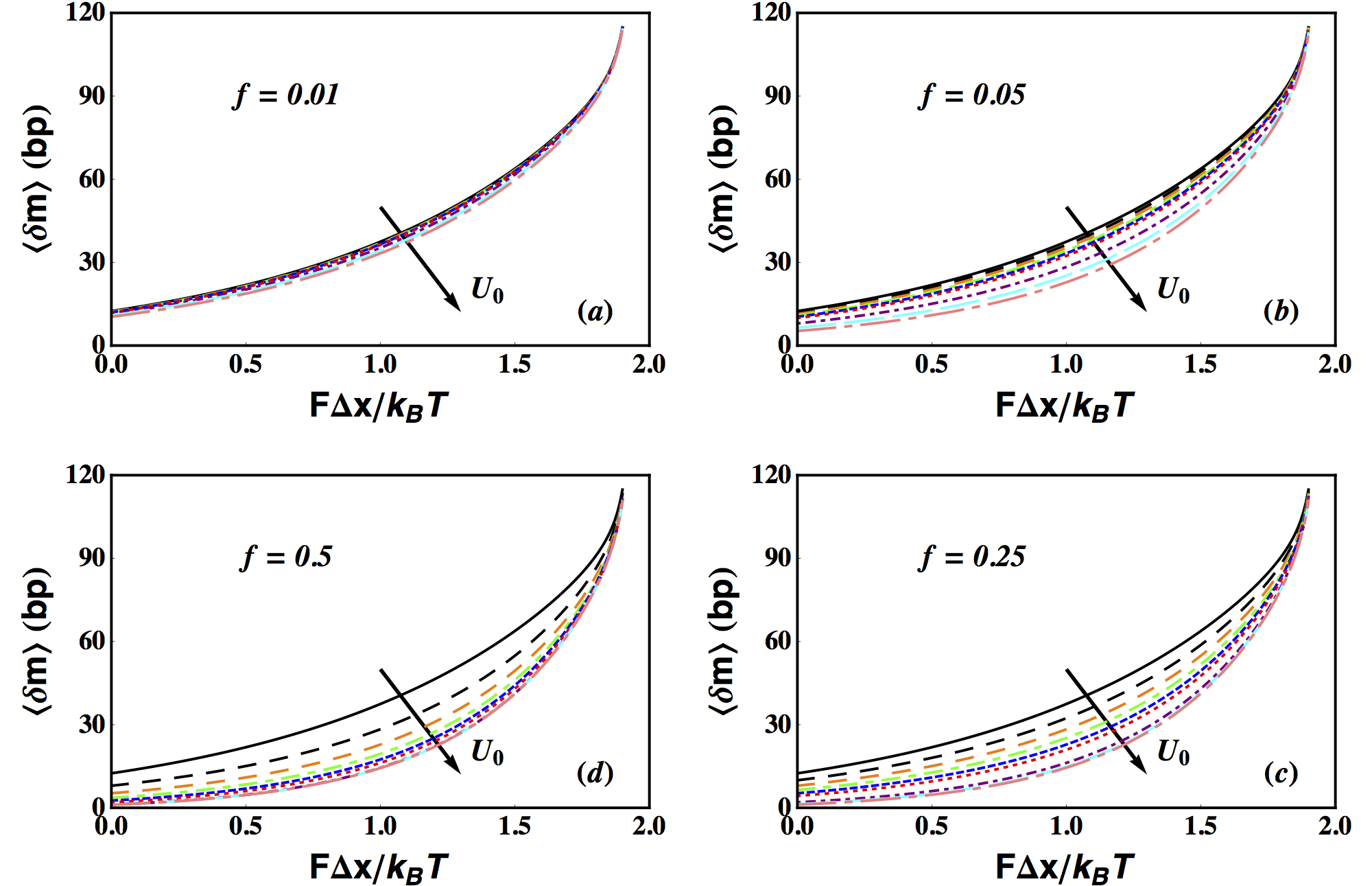}
\end{center}
\caption{Plots of the mean unwinding processivity of a helicase as a function of the applied tension, for $U_{0}/k_{B}T=$0 (passive),1, 2, 3, 4, 5, 10, 15, and 20.\ \ The processivity always
increases with increasing tension destabilizing the ss-dsNA junction
and decreases with increasing step height.\ \ We used the same parameters as in Fig.~\ref{XRef-FigureCaption-422164312}.}
\label{up}
\end{figure}

\item  The unwinding processivity depends strongly on the parameter $f$.\ \ For $f=0.01$ and
$F=0$, $\langle \delta m\rangle /{\langle \delta m\rangle }_{\mathrm{HW}}>0.8$
even when the coupling potential $U_{0}=20 k_{B}T$ (pink dot-dashed
line in Fig.\ \ref{XRef-FigureCaption-422164312}).\ \ For $f=0.05$
and $F=0$, the coupling potential $U_{0}$ must exceed $5 k_{B}T$
for $\langle \delta m\rangle /{\langle \delta m\rangle }_{\mathrm{HW}}$ to
decrease to less than 0.8 (red dotted curve in Fig.\ \ref{XRef-FigureCaption-422164312}).\ \ Thus,
the effects of the barriers to translocation and NA breathing on
$\langle \delta m\rangle $ are the principal determinants of
the `velocity' of a processive helicase (see also sequence effects below).
\item The equation $\left\langle  \delta m\right\rangle= V_1(F,U_0) \left\langle  \tau \right\rangle$ gives key insights into the force dependence of processivity. As was discussed above, for a passive helicase with $U_0=0$, the velocity $V_{\text{HW}}$ increases rapidly as the force increases (Fig.~\ref{XRef-Figure-422171440}). However, the lifetime of a passive helicase $\langle  \tau \rangle_{\text{HW}}$ does not change and is constant at $\langle  \tau \rangle_{\text{HW}}=1/\gamma$ (Eq.~\ref{tauhw}). This immediately means that $\langle  \delta m \rangle$ will increase rapidly with increase in force (solid black lines in Fig.~\ref{up}). In the other extreme limit of an optimally active helicase ($U_0/k_B T \gg 1$ and $f \ll 1$), our discussions above show that $V_1$ will not significantly change with force (Fig.~\ref{XRef-Figure-422171440}). However, unlike the passive situation, $\langle  \tau \rangle$ increases rapidly with force (Fig.~\ref{XRef-FigureCaption-422164254}). This result follows again from the discussion in the previous points---as a helicase becomes more active, the increased interaction with the junction reduces its lifetime. An external force reduces the probability of this interaction which in turn results in an increase in $\langle  \tau \rangle$, the lifetime of the helicase. As a result, $\langle  \delta m \rangle$  will increase rapidly as the external force increases (Fig.~\ref{up}a). For intermediate values of $U_0$ and higher values of $f$, the increase in $\langle  \delta m \rangle$ with force is a result of contributions from both the velocity as well as the lifetime. Thus, the phenomenon of rapid increase in unwinding processivity as $F$ increases, arises due to very different reasons, depending on how active the helicase is. These arguments lead to the surprising prediction that no matter how active (or passive) the helicase is, the processivity is sensitive to external force. Our prediction, which we believe is universal for unwinding helicases,  is borne out in the few experiments that have analyzed the variation of unwinding processivity over a range of forces \cite{johnson_single-molecule_2007, DumontNature06}. This behavior is to be contrasted with the dependence of unwinding velocity on $F$, which varies significantly when the helicase is passive, but less so for active helicases. 
\end{enumerate}

\emph{Sequence effects}.  The quantitative insights obtained for helicase velocity  (Fig.\
\ref{XRef-Figure-422171440}) and processivity (Fig.~\ref{up}) as a function of force  prompted us to use a similar model
to explore the effects of NA sequence.  Helicase unwinding and translocation
can be modeled as a discrete-state continuous-time stochastic process \cite{kampen_stochastic_2007}. Instead 
of adopting  Eq.\ \ref{XRef-Equation-419155226} to include sequence effects, we used the 
Kinetic Monte Carlo (KMC) method \cite{Newman2004} to simulate the model. We chose KMC in this case because grid sizes 
larger than $M=10^4$ would be needed to solve  Eq.\ \ref{XRef-Equation-419155226}, which contributes to numerical stability problems.
The transitions in the KMC are stochastically implemented, and a given trajectory is generated till the helicase disassociates.

From detailed balance it follows that $\frac{\alpha }{\beta }= e ^{-\Delta
G}$, where $\Delta G\approx 2$ is the sequence averaged free-energy
per base pair in the absence of the helicase.\ \ To incorporate sequence
effects we let $\Delta G=g( \delta m) $, where $\delta m\equiv m-m_{0}$,
$m_{0}$ is the initial position of the ss-dsNA junction on the lattice
and the sequence-dependent function $g$ is constructed from the nearest-neighbor parameters provided in
Table 3 of SantaLucia \textit{et al}.\ \cite{santalucia_improved_1996}.\ \ For example,
if the junction is a GC base pair and the downstream pair 
is AT we assign $\Delta G( \delta m) =1.46$ $\mathrm{ kcal}/\mathrm{mol}\approx
2.43 k_{B}T$.\ \ Following BJ \cite{betterton_velocity_2005}, we assume
$\beta $ to be constant and then assign $\alpha =\beta  e ^{-\Delta
G}$.\ \ As the sequence composition is varied, the average free energy
per base pair ranges from $\overline{\Delta G}=1.7$ when the GC content is 0\%
to $\overline{\Delta G}=2.95$ if the GC content is 100\%. We investigated the NA
sequences organized in a block copolymer fashion.\ \ For
example, the infinitely repeating unit of the sequence used to investigate
$40\%$ GC content was $\begin{array}{cccccccccccc}
 5^{\prime } & G & G & G & G & A & A & A & A & A & A & 3^{\prime
} \\
 3^{\prime } & C & C & C & C & T & T & T & T & T & T & 5^{\prime
}
\end{array}$.\ \ We investigated sequences with fractional GC content
varying from 0 to 1 in increments of 0.1.\ \ All simulations were
performed with initial separation $j_{0}=1$ and with the junction
located at the first GC pair of the repeating unit of the sequence.

The mean attachment time is calculated as:
$\left\langle  \tau \right\rangle  \equiv 1/N\sum \limits_{i=1}^{N}\tau
_{i}$, where $N$ is the number of simulations ($N=1000$ for every
data point we collected), and $\tau _{i}$ is the attachment time in
simulation $i$.\ \ The mean translocation processivity is calculated using 
$\left\langle  \delta n\right\rangle  \equiv 1/N\sum \limits_{i=1}^{N}\left(
n_{i}^{f}-n_{i}^{0}\right)$, where $n_{i}^{0}$ is the initial position of the helicase is simulation $i$ and $n_{i}^{f}$ is its final position.\ \ Finally,
the mean unwinding processivity is calculated as $\left\langle  \delta m\right\rangle  \equiv 1/N\sum \limits_{i=1}^{N}\left(m_{i}^{f}-m_{i}^{0}\right)$, 
where $m_{i}^{0}$ is the initial position of the ss-dsNA junction and $m_{i}^{f}$ is its final position.

The results of our simulations investigating sequence dependence are provided in Fig.\ \ref{XRef-FigureCaption-422164915}, and we note the following points of interest:
(1)  Not surprisingly, processivity (both $\langle \delta m\rangle
$ and $\langle \tau \rangle $) and $V_1=\langle \delta m\rangle /\langle
\tau \rangle $ decrease as the fraction of GC content is increased.
(2)  The decrease in attachment time with increasing $U_{0}$ is
similar at both $f=0.01$ and $f=0.25$.\ \ The behavior of $\langle
\delta m\rangle $ with increasing $U_{0}$ at $f=0.25$, however, differs
substantially from its behavior at $f=0.01$.\ \ At $f=0.01$ $\langle
\delta m\rangle $ is essentially constant with respect to $U_{0}$.\ \ This
leads to an eventual saturation of $\langle \delta m\rangle /\langle
\tau \rangle $ at large $U_{0}$.\ \ At $f=0.25$, on the other hand,
sequence effects are more pronounced.\ \ For example, $\langle \delta
m\rangle $ decreases substantially with $U_{0}$ when the $\%\mathrm{GC}=0$ (Fig.\ \ref{XRef-FigureCaption-422164915}, black
solid line), but negligibly
when $\%\mathrm{GC}=1$ (Fig.\ \ref{XRef-FigureCaption-422164915}, grey dashed line).\ \ This
behavior leads to a pronounced peak in $V_1$ when $\%\mathrm{GC}=0$, but a very small
peak when the fractional GC content is 1.\ \ Indeed, at $\%\mathrm{GC}=100$,
we see that $\langle \delta m\rangle $ can actually become negative
leading to a negative $V_1$ for helicases with too strong a coupling. {\bf The surprising finding that  velocity and processivity can be negative can be explained as follows.  If the base ahead of the helicase is very stable, with a large $\Delta G$, and if the helicase back-stepping rate (during unwinding) is comparable in magnitude to the forward stepping rate, it is possible that on average the helicase steps backward before the base in front opens from a thermal fluctuations. In such a situation, the velocity and processivity would be negative. It remains to be ascertained if this is realized in practice.}

\section{Discussion}
\begin{figure}
\begin{center}
\includegraphics[width=8cm]{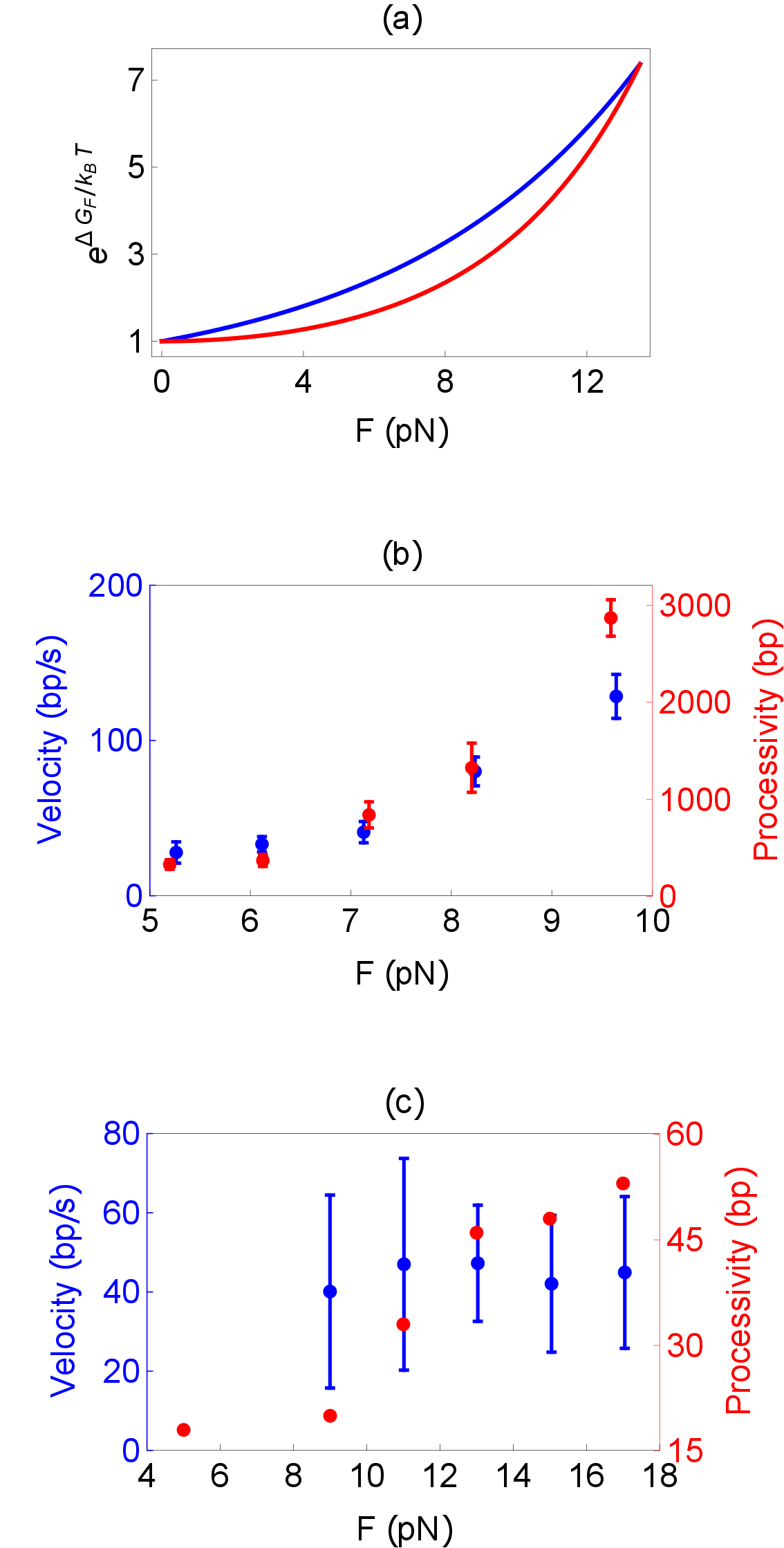}
\end{center}
\caption{(a) Dependence of $e^{\Delta G_F}$ on $F$ for two models: $\Delta G_F=F \Delta x$ (blue curve), $\Delta G_F=2\frac{L}{l}\text{log}\left(\frac{1}{F l}\text{sinh}(F l)\right)$ (red curve). The parameters $\Delta x=0.594$ nm, $L=0.6$ nm/nucleotide and $l=1.3$ nm were chosen such that for both models, the critical force (force at which $\Delta G=\Delta G_F$) is 13.5 pN, a typical value for DNA hairpins. To be consistent with the rest of our analysis, $\Delta G$ was chosen to be $1.95 \, k_BT$. (b) and (c) Experimental data suggesting a universal behavior of the unwinding processivity as a function of force. Velocity (blue) and processivity (red) data on (b) the T7 helicase \cite{johnson_single-molecule_2007} and (c) the NS3 helicase \cite{DumontNature06}. The data shows that the unwinding velocity of the two helicases can be strongly or weakly dependent on external force. The processivity clearly increases as F increases, for both the helicases.}
\label{universal}
\end{figure}

\textit{Effect of ssDNA elasticity:} To simplify our analysis, we used the Bell model $\Delta G_F=F \Delta x$, for the effect of force on the destabilization of the junction base-pair. It might be more accurate to use $\Delta G_F=2\frac{L}{l}\text{log}\left(\frac{1}{F l}\text{sinh}(F l)\right)$ \cite{cocco_force_2001, lionnet_real-time_2007}, because a freely-jointed-chain model has been shown to be appropriate to describe single strand DNA elasticity \cite{toan_2010}. As shown in Fig.~\ref{universal}a, $e^{\Delta G_F}$ for both the models are similar, and hence using the simpler Bell model does not make any qualitative difference in our results.

\textit{Effect of back-stepping rate:} We analyzed the model for the case where the back-stepping rate of the helicase $k^{-}$ is much smaller than the forward stepping rate $k^+$, in the absence of the double-strand junction. Most molecular motors fall in this regime, where $k^+ \gg k^-$. For example, the ratio $k^+/k^-$ was measured to be about 221 in Kinesin \cite{Nishiyama2002} and is expected to be large for helicases as well \cite{manosas_active_2010}. As long as this holds good, which has experimental support, all the results obtained in this work will remain valid. 

\textit{Mechanism for increase in processivity with force is different for passive and active helicases:} From the recent single-molecule experiments on three helicases \cite{DessingesPNAS04,DumontNature06,johnson_single-molecule_2007}, we have surmised that the dependence of unwinding processivity on $F$ may be universal (Fig.~\ref{universal}b and c ).  Irrespective of whether the unwinding velocity increases rapidly with force or not, the processivity seems to increase with the external force. In this work, we have provided a theoretical explanation of this behavior. Based on the theory, we predict that the sensitivity of processivity to force should indeed be a universal feature of all helicases, whether they are active or passive. 
Our argument hinges on the observation that the processivity of a helicase is very well approximated by the product of two quantities, the unwinding velocity and the attachment time of the helicase: $\left\langle  \delta m\right\rangle \approx V_1 \left\langle  \tau \right\rangle$. The origin of the universal increase of $\left\langle  \delta m\right\rangle $  with $F$ is dramatically different for passive and active helicases. We have shown that when the helicase is passive, $V_1 (=V_{\text{HW}}$) increases rapidly with force while $\left\langle  \tau \right\rangle_{\text{HW}}$ stays constant, independent of the force. In contrast, when the helicase is optimally active, $V_1$ hardly changes as a function of force while $\left\langle  \tau \right\rangle$ increases rapidly as the force is ramped up. Thus, in both these extreme situations of helicase activity, the processivity $\left\langle  \delta m \right\rangle$ shows significant variation as a function of force. This leads to the prediction that irrespective of the nature of interaction of the helicase with the ds junction, the processivity should always increase as the force is increased.

{\bf We note {\it en passant} the relationship, $\left\langle  \delta m\right\rangle \approx V_1 \left\langle  \tau \right\rangle$, can be used to anticipate the effect of ATP concentration on $\left\langle  \delta m \right\rangle$. If for a particular helicase $\left\langle  \tau \right\rangle$ is independent of ATP concentration, it follows that processivity is determined by $V_1$. So we expect that decreasing ATP concentration should decrease $V_1$, and hence $\left\langle  \delta m \right\rangle$. Interestingly, this was observed in an experiment on RecBCD sometime ago \cite{roman_characterization_1989}}. 

\textit{Is there a general role for partner proteins in the replisome?:} Our demonstration that helicase processivity always increases with $F$ regardless of the underlying architecture of the enzyme, is likely to be relevant \textit{in vivo} as well. It is becoming increasingly evident that helicases function most efficiently only in the presence of the macro-molecular machinery with which it interacts in the cellular milieu \cite{von_hippel2001,lohman_non2008}. Significant increase in the unwinding velocity and processivity of a number of helicases has been reported \cite{stano2005,rajagopal_single_2008,manosas_collaborative_2012}, when coupled to partner proteins like the polymerase or single-stranded binding proteins (SSB). This naturally suggests a possibility that by partially destabilizing the ss-dsDNA junction, the partner proteins could be mimicking external forces that are applied in single molecule experiments. 

A number of explanations have been proposed in the literature, to account for the effect of partner proteins on helicases \cite{von_hippel2001}. For example, dsDNA destabilization, prevention of back-slippage of the helicase, specific interactions with the helicase leading to longer helicase lifetimes on the nucleic acid track, are some of the commonly invoked methods by which partner proteins are conjectured to increase the efficiency of helicases. Though it is likely that some or all of these mechanisms together contribute in increasing a helicase's speed and processivity, there is evidence to suggest that partial destabilzation of the ss-dsDNA junction may be essential. It has been known for a long time that by binding tightly to ssDNA, SSB can destabilize the bases of dsDNA \cite{meyer_single1990,witte_single_2005}, at least the first few bases near the ss-ds junction \cite{eggington_polar_2006}. A similar double-strand destabilization effect caused by the polymerase was shown to be critical in explaining single-molecule data on the T4 helicase \cite{manosas_collaborative_2012}. To explain the sometimes order-of-magnitude increase in velocity/processivity of helicases, it therefore seems likely that an essential requirement is at least partial destabilization of a few base pairs at the ss-dsDNA junction, by partner proteins. 

 If partner proteins indeed mimic external forces resulting in destabilization of the ss-dsDNA junction, we would predict that they would have distinctly different effects on T7 (weakly active) and NS3 (active) helicases. The effects should be similar to the force response (Fig.~\ref{universal} b,c) -- both the velocity and processivity of T7 helicase should be increased by partner proteins, while \textit{only} the processivity should be enhanced of the NS3 helicase. Remarkably, this has indeed been observed in experiments. Rajagopal and Patel \cite{rajagopal_single_2008} showed that SSB increase only the processivity of NS3 while Stano \textit{et al} \cite{stano2005} and previous works \cite{notarnicola_1997} showed that both velocity and processivity of T7 helicase increases due to the presence of a polymerase.
 
\textbf{It is worth re-emphasizing that we are drawing a parallel between the effect of partner proteins on the ss-dsDNA junction, and the base-pair destabilization effect (reduction in $\Delta G$) due to  external mechanical forces. We are not predicting the effect of partner proteins on the individual rates whose ratio determine $\Delta G$ (see Eq.~\ref{XRef-Equation-416145310}). Therefore, destabilization of the junction by partner proteins could happen either by increasing the dsDNA opening rate $\alpha$ or by decreasing the dsDNA closing rate $\beta$. The latter mechanism was shown to hold in a particular study of the effect of single strand binding proteins on dsDNA \cite{hatch_direct_2007}, and hence is another example supporting our claim.}  

Our work therefore suggests a very general role for the partner proteins associated with helicases in the replisome. Even though these associated proteins may or may not increase the unwinding velocity of a helicase depending on the specifics of the helicase's interaction with the dsDNA, they should universally increase the processivity of the helicase. This seems to be borne out in experiments, and the generality of our statement can be tested in future experiments. \textbf{In a related context, it was shown earlier that the interactions between separate domains of the same helicase could potentially lead to a higher velocity than either individual domain \cite{evgeny_coupling_2005}. This mechanism could also play a role in understanding the reason for increase in velocity of different oligomeric states of helicases or helicases interacting with partner proteins.}

\section{Conclusions}
There has been no unified explanation for the recent experimental observations that the unwinding processivity of helicases is always enhanced by force while the velocity shows a multitude of helicase-dependent responses. By extending the theory of Betterton and J\"ulicher to include force effects, we have provided an understanding of these varied observations. The velocity of unwinding depends crucially on the extent to which a helicase is active. Optimally active helicases show little or no change in velocity while the velocity of passive helicases is highly sensitive to external forces that destabilize the double strand. In stark contrast, we predict that the unwinding processivity of a helicase should always increase with force, irrespective of how active or passive it is. The reason for this universal behavior of the processivity however, depends on the nature of the helicase. Our prediction is general, and seems to be borne out in structurally diverse helicases like T7, UvrD and NS3.  Future experimental measurements of load-dependent processivity can test our assertion that associated partner proteins may have coevolved with helicases to increase the processivity, not the velocity of unwinding nucleic acids. Finally, our sequence dependent simulations show that the velocity
can display non-monotonic behavior that is strongly dependent on GC content as well as saturating behavior at very small $f$. Many of our predictions await future experiments.

\vspace{10 mm}

\textbf{Author Contributions}: DP, SC, and DT designed and performed research and wrote the paper.

\vspace{10 mm}

\textbf{Acknowledgements}: We are grateful to Professor Stephen Kowalczykowski and Professor Tim Lohman for their interest and for useful comments. This work was supported in part by a grant from the National Science Foundation (CHE13-61946).


\end{document}